\documentclass[preprint2]{aastex}

\shorttitle{The Solar System as a Dark Matter Probe}
\shortauthors{X. Hernandez}

\begin{document}


\title{Detailed Solar System dynamics as a probe of the Dark Matter hypothesis}


\author{X. Hernandez}
\affil{Instituto de Astronom\'{\i}a, Universidad Nacional Aut\'{o}noma de M\'{e}xico, Apartado Postal 70--264 C.P. 
04510 M\'exico D.F. M\'exico.}



\begin{abstract}
  Within the dark matter paradigm, explaining observed orbital dynamics at galactic level through the inclusion
  of a dominant dark halo, implies also the necessary appearance of dynamical friction effects. Satellite galaxies,
  globular clusters and even stars orbiting within these galactic halos, will perturb the equilibrium orbits
  of dark matter particles encountered, to produce a resulting trailing wake of slightly enhanced dark matter
  density associated with any perturber in the halo. The principal effect of this gravitational interaction between
  an orbiting body and the dark matter particles composing it, is the appearance of a frictional drag force slowly
  removing energy and angular momentum from the perturber. Whilst this effect might be relevant to help bring about
  the actual merger of the components of interacting forming galaxies, at smaller stellar scales, it becomes negligible.
  However, the trailing wake will still be present. In this letter I show that the corresponding dark matter wake
  associated to the Sun, will constitute a small but resonant perturbation on solar system dynamics which can be ruled out,
  as current radio ranging measurements are now {close to} an order of magnitude more precise than the amplitude of the
  orbital perturbations which said wake implies. The absence of any such detection implies the nonexistence of the dynamical
  friction trailing wake on the sun, which in turn strongly disfavours dark matter as an explanation for the observed
  gravitational anomalies at galactic scales.
\end{abstract}



\section{Introduction}

Over the past years the number and variety of proposals offered as alternative explanations to the dark matter hypothesis as
solutions to the well established gravitational anomalies at galactic and cosmological scales, has significantly increased.
Starting with MOND (Milgrom 1983) and MOG (Moffat 2005), the list now includes the covariant schemes of conformal gravity
(Mannheim 2006), f(R) extensions to GR (e.g. Capozziello et al. 2007 or Barrientos \& Mendoza 2018), the quantised inertia
proposal of McCulloch (2012) the emergent gravity holographic ideas of Verlinde (2017), the mimetic models of Vagnozzi (2017),
the modified inertia model of Van Putten (2018) and the exploration of negative mass scenarios recently developed by Farnes (2018).
While the preceding list is not exhaustive, it illustrates the increasing amount and variety of work going into developing ideas
alternative to the dark matter paradigm. This has been partly in response to the various problems which have arisen with the
standard picture in terms of details (e.g. Famaey \& McGaugh 2012), but mostly, due to the continual and complete lack of a direct
detection or an independent confirmation of the existence of dark matter, beyond gravitational anomalies in the low acceleration regime.

Recently we have seen all direct detection experiments return only null results (e.g. Yang 2016), as sensitivity limits previously
deemed important have been reached and surpassed. The same is the case for astronomical searches for hypothetical dark matter
self-annihilation signals (e.g. {\it Fermi}-LAT Collaboration 2016). The current dark matter paradigm is a framework where
the driving causal entity is something which no one has ever seen a single particle of.

Whilst alternative scenarios are constructed primarily to reproduce the dynamics otherwise attributed to the dark component,
a second-order effect of these hypothetical particles offers an independent window into testing their reality.
A dwarf galaxy, star cluster, or even a single star, while moving through a hypothetical dark matter halo, will not only react
to the total potential of the overall dark matter distribution, but also interact and deflect each single dark matter particle
it encounters along its orbit. The collective effect of all such interactions will gradually remove energy and angular momentum
from the orbiting body through a process known as dynamical friction. As an unavoidable consequence of this dynamical friction,
a trailing wake of dark matter will form behind the orbiting body.

Indeed, dynamical friction is an integral part of the dark matter paradigm, responsible for making merger timescales of
the constituents of forming galaxies compatible with observational constraints on galactic formation scenarios, e.g.
White (1976). Dynamical friction considerations have often been used to constrain dark matter properties, such as dark halo
density profiles through requiring the survival of observed globular clusters in dwarf spheroidal galaxies (e.g. Hernandez \&
Gilmore 1998, Goerdt et al. 2006), the consistency of the observed morphology of shell galaxies in dark matter models
(Vakili et al. 2017), the merger dynamics of forming galaxies (Pe\~{n}arrubia et al. 2002,  Kroupa 2015) or the survival of
ultra-faint dwarf galaxies (Hernandez 2016). {In terms of future tests, Pani (2015) showed that dynamical friction on binary
pulsars detected close to the galactic centre could also yield interesting restrictions.}

Extending such ideas, in this letter I show that the dark matter dynamical friction wake which should trail the sun, constitutes an
essentially resonant perturbation on the orbital dynamics of the solar system. In spite of such resonant character, the relative
radial amplitude of the expected perturbation on the planets is of only $\sim 10^{-10}$. This implies radial amplitudes
of only 11 and 252 meters for the cases of Mars and Saturn, respectively. However, over the past decades,
radio tracking experiments to Mars and Saturn have reached resolutions of centimetres and meters respectively. This has allowed the
comparison of observations to theoretical ephemerides, calculated now through very high order integration of Solar System models
including the perturbations of all the planets, over 300 of the largest asteroids, several tens of trans-Neptunian objects and the
asteroid belt including post-Newtonian approximations and the effects of non-sphericity and extended body mutual interactions e.g.
Folkner et al. (2014), Viswanathan et al. (2017). Residuals of these models when compared to observations are now of $0.06m \pm 3.6m$
and $5m \pm 32.1m$ for the cases of Mars and Saturn, respectively e.g. the latest INPOP17a ephemerides, Viswanathan et al. (2017).

The absence of any unaccounted for orbital deviations at scales {close to} an order of magnitude above current model-observations residuals,
after all Solar System perturbations are included, implies that the aforementioned dark matter wake is not present. This
in turn strongly favours extended gravity proposals as explanations for the observed gravitational anomalies at galactic
scales.

\section{Analysis of the response to the Solar Dark Matter wake}

I begin with the description of the dynamical friction wake trailing a point mass $M$ travelling at the centrifugal
equilibrium velocity $V$ within a dark matter halo having a Maxwellian velocity distribution function
with dispersion $\sigma$ and a dark matter density at the location of the perturber $\rho_{0}$ (Mulder 1983, Binney \&
Tremaine 2008):

\begin{equation}
\rho_{w}=\rho_{0} \frac{G M}{\sigma^{2} r} e^{-sin^{2}\theta} \left[1-erf(cos\theta) \right],
\end{equation}


\noindent where I have taken $V=2^{1/2} \sigma$, $r$ is the radial distance to the perturber, and $\theta$ a polar
angle such that $\theta =\pi$ points in the direction of motion of the perturber. The presence of this wake implies
a force pointing away from the direction of travel, $F_{DF}$ {which constitutes dynamical friction. In the context of calculating
the perturbation to planetary orbits which this force induces, we are interested in the residual acceleration on a planet,
after subtracting the acceleration felt by the sun. This constitutes what is termed the deforming force,
$F_{def}(r, \theta)=F_{DF}(r, \theta)-F_{DF}(r=0)$, and which along the axis of symmetry is given by (Mulder 1983): }

\begin{equation}
F_{def}(r)=4 \pi \frac{G^{2} \rho_{0} M_{\odot}}{\sigma^{2}} \left[0.21 ln(R_{p}/2r_{min})+0.44 \frac{cos \theta}{|cos \theta|} \right].
\end{equation}

\noindent In the above I have taken $M=M_{\odot}$ to estimate the perturbing force due to the dynamical friction wake
which the Sun would excite if travelling through a dark matter halo.
{This becomes a compression force in the trailing direction, and a force which pulls away in the leading one, which in the
frame of reference of the sun can be modelled as having a $cos(\Omega_{w}\theta)$ angular dependence,  with $\Omega_{w}$ the frequency
of the dynamical friction wake, given the close to anti-symmetric character of equation (2). Notice that this residual force has
no dependence on the total size of the system, the usual $R_{Max}$ parameter of dynamical friction, as it is a residual dependent
only on local conditions of the wake. The parameter $r_{min}$ is the radius at which the circular velocity of a planet becomes equal
to the Galactic orbital velocity of the sun, so that $2r_{min}=0.037 AU$.}

The very mild logarithmic dependence of the above force on the radial separation between the Sun and the planet
being considered, guarantees that the potential of this wake can be taken as

\begin{equation}
\Phi_{w}=-r F_{def}(r).
\end{equation}

{We can now estimate the response of a planet to the above perturbing acceleration} through
a perturbative analysis of their orbits, in the presence of a $m=1$ symmetric perturbation, analogous to the case of $m=2$
bars perturbing galactic stellar orbits or $m=4$ for a 4-armed spiral, e.g. the development leading to eq. 3.118a in
Binney \& Tremaine (1987), which in the present context yields:

\begin{equation}
  \ddot{R_{1}}+\kappa^{2}R_{1}=-\left[\frac{d\Phi_{w}}{dr} +\frac{2 \Phi_{w}\Omega_{0}}{r(\Omega_{0}-\Omega_{w})}
    \right]_{R_{p}}cos[t(\Omega_{0}-\Omega_{w})],
\end{equation}

\noindent for the radial perturbation, $R_{1}$, on the orbit of a planet at radius $R_{p}$ due to the presence of the mass distribution
of eq.(1) producing a wake potential $\Phi_{w}$. In the above $\kappa$ and $\Omega_{0}$ are the epicycle and
orbital frequencies of a planet being considered. This equation is just the simple harmonic motion of the epicycle
approximation, in addition to which a forcing term given by the wake potential $\Phi_{w}$, and having a frequency
$(\Omega_{0}-\Omega_{w})$ appears. Notice that $\Omega_{w}$ is the frequency of the dark matter wake, which enters as the
analysis leading to equation (4) has been performed in a rotating frame in which the dark matter wake is static. This
dark matter wake points away from the solar motion at all times, and hence, in the course of the solar orbit around the
Galaxy, will rotate with respect to the solar system, and also shift in and out of the solar system plane, with a frequency
given by $\Omega_{w}=2 \pi/ T_{\odot}$, where $T_{\odot}=2.5 \times 10^{8}$ yr is the orbital period of the Sun about the Galaxy.
The amplitude of the forced mode is now:

\begin{equation}
  \Delta R_{1}=\frac{-1}{\kappa^{2}-(\Omega_{0}-\Omega_{w})^{2}}\left[\frac{d\Phi_{w}}{dr} +\frac{2 \Omega_{0}\Phi_{w}}
  {r(\Omega_{0}-\Omega_{w})}  \right]_{R_{p}}
\end{equation}

Since $\kappa^{2}=r d\Omega_{0}^{2}/dr +4\Omega_{0}^{2}$, for kepplerian motion we have $\kappa=\Omega_{0}$.
From the above equation we see that the amplitude of the response to the dynamical friction wake becomes
resonant and formally diverges as $\Omega_{w} \rightarrow 0$. Since $\Omega_{0}=2\pi/T_{p}$ and
$\Omega_{w}=2\pi/T_{\odot}$, and given that $T_{\odot}>>T_{p}$  for the orbital period of any Solar System planet
$T_{p}$, we can write the amplitude of the quasi-resonant mode as:

\begin{equation}
\Delta R_{1}= \frac{-1}{2\Omega_{0}\Omega_{w}}  \left[\frac{d\Phi_{w}}{dr} +\frac{2 \Phi_{w}}{r}  \right]_{R_{p}}
\end{equation}

Using equation(3), the term in brackets above becomes $-3F_{def}$. 
Taking $\Omega_{0}=(G M_{\odot})^{1/2}/R_{p}^{3/2}$, the relative amplitude of the quasi-resonant mode for a
planet orbiting at a distance $R_{p}$ from the sun becomes:

\begin{equation}
\frac{\Delta R_{p}}{R_{p}}= \pi \frac{G^{3/2}\rho_{0} M_{\odot}^{1/2} R_{p}^{1/2}}{\sigma^{2} \Omega_{w}}  \left[0.21 ln(R_{p}/0.037 AU)+0.44\right] 
\end{equation}

\noindent where we have taken an angle
between the plane of the Solar System and the Sun's direction of travel of $60$ degrees, when calculating the above radial
perturbing force on the plane of the Solar System. Taking $\rho_{0}=0.01 M_{\odot} pc^{-3}$ (e.g. Read 2014), $\sigma=220 km
s^{-1}/2^{1/2}=156 km s^{-1}$ and $T_{\odot}=2.5\times10^{8} yr$, we obtain:
{
\begin{eqnarray}
  \frac{\Delta R_{p}}{R_{p}} =& 3.24 \times 10^{-11} \left[0.21 ln(R_{p}/0.037 AU) \right. \nonumber \\
    & + \left. 0.44\right]  \left(\frac{R_{p}}{AU} \right)^{1/2},
\end{eqnarray}

\noindent which is the main result of this section. 
}

\section{Comparisons with observational limits}

The results of the previous section can now be compared to the residuals of comparing radio tracking of spacecraft landed or in orbit
about Saturn and Mars. As described in the
introduction, the latter now include all known perturbers in the Solar System, extended body and relativistic corrections. Residuals in orbital
distances of the latest INPOP17a ephemerides to radio tracking experiments are now of $\sim$ cm residuals with uncertainties of 3.6 meters
in the case of Mars, and 5.1 meter residuals with uncertainties of 31.6 meters in the case of Saturn, Viswanathan et al. (2017).

{Using equation (8) for the case of Mars, $R_{p}=1.52 AU$, we get an amplitude due to the perturbing acceleration calculated
of 11.26 meters. This is 3.13 times larger than the confidence interval of the essentially zero residuals of the latest ephemerides, a result
which is expected with a probability of only 0.17\%, should the perturbing dark matter wake actually be there. For the case of Saturn
with $R_{p}=10 AU$, the amplitude of the expected response from equation (8) now becomes 252 meters, 7.85 times larger than
the confidence intervals of the most recent celestial dynamics calculations, consistent with the expected dark matter wake with
a probability of only $5.7\times 10^{-15}$.}

Given the linear relation between the amplitude of the expected orbital perturbations and the assumed dark matter density of
equations (2) and (7), within a dark matter scenario, the above results become upper limit restrictions on the local density of
dark matter at the position of the Sun, of below 3.13 and 6.8 times (for Mars and Saturn, respectively) the
preferred values of $0.01 M_{\odot} pc^{-3}$, see Read (2014), to force the expected response to lie within the reported confidence intervals
of current ephemerides. It is interesting that dynamical constraints from requiring
agreement between the observed rotation curve of the Milky Way and dark matter halo models, yield also a fairly tight interval of
$\rho_{0}=0.01 \pm 0.005 M_{\odot} pc^{-3}$, e.g. Iocco et al. (2011) and references therein, a variation of only a factor of 2. Thus, the
limits derived in this letter show that assuming a dark matter halo to explain the orbital dynamics of the Milky Way is inconsistent
with observations of Solar System dynamics not showing any detectable sign of the radial perturbations which the dark matter wake
associated to the assumed halo would elicit.

The use of Planetary Solar System celestial mechanics as constraints on the dark matter hypothesis follows the work of e.g. Pitjeva
\& Pitjev (2013), who by considering only a spherically symmetric dark matter component within the Solar System, and taking
observational limits on any deviations from keplerian motion for the Planets, set limits of $\rho_{0}<130 M_{\odot}pc^{-3}$ for the
local dark matter density. This limits are about 5 orders of magnitude less restrictive than what I derive in this letter, considering
the much more disruptive asymmetric dynamical friction quasi-resonant wake. 

A caveat to the accuracy of the orbital perturbations calculated here comes from the fact that I have only considered the component
of the force due to the dynamical friction wake, in the plane of the Solar System. A similar component would appear out of the
plane, having the effect of producing vertical perturbations, which taken together with the radial ones considered here, would
result in even tighter bounds. A fuller analysis must consider the shifting position of the dark matter wake over the orbital
period of the Sun around the Milky Way, integrating over periods of order $T_{\odot}$ or larger. This extended analysis should also
consider all planets simultaneously, and will probably yield similarly interesting constraints on the allowed local dark matter
density, which might have a bearing on the long term stability of the Solar System taken in its entirety. 

The argument presented applies equally to any dark matter particle candidate, as the gravitational interaction between axions,
wimps or machos and the Sun would be the same. Indeed, even more exotic scalar field/fuzzy dark matter (e.g. Matos et al. 2000)
or superfluid dark matter proposals, which behave as normal dark matter in the vicinity of the Sun, (e.g. Berezhiani et al. 2018)
would result in dynamical friction (as required to yield consistent structure formation scenarios), although the details will
to some degree vary e.g. Du et al. (2017).

\section{Conclusions}  

In summary, models which extend the current general relativity theoretical scheme, eliminating the need for a dominant halo of
gravitationally interacting particles to account for galactic dynamics, appear generaically more feasible than dark matter
proposals resulting in a dynamical friction wake trailing the Sun, which due to the non-detection of the effects such asymmetric
density distribution would entail, can be discarded.

\section*{acknowledgements}

Xavier Hernandez acknowledges financial assistance from UNAM DGAPA grant IN104517 and CONACYT.

\end{document}